\documentclass[12pt,preprint]{aastex}
\usepackage{natbib} 

\usepackage{amsmath} 
\usepackage{pdflscape}
\usepackage{multirow}
\usepackage{makecell}

\usepackage{color}

\begin{document}
\shorttitle{A Wavelet Based Foreground Subtraction}
\shortauthors{Gu et al.}

\title{The Application of Continuous Wavelet Transform Based
  Foreground Subtraction Method in 21 cm Sky Surveys}

\author{Junhua Gu\altaffilmark{1}, Haiguang Xu \altaffilmark{2},
  Jingying Wang \altaffilmark{2}, Tao An \altaffilmark{3,4}, and Wen
  Chen\altaffilmark{2}}

\altaffiltext{1}{National Astronomical Observatories CAS, 20A Datun
  Road, Chaoyang District, Beijing 100012, China; jhgu@bao.ac.cn}
\altaffiltext{2}{Shanghai Jiao Tong University, 800 Dongchuan Road,
  Shanghai 200240, China} \altaffiltext{3}{Shanghai Astronomical
  Observatory CAS, 80 Nandan Road, 200030, Shanghai China}
\altaffiltext{4}{Key Laboratory of Radio Astronomy, Chinese Academy of
  Sciences, 210008, Nanjing, China}

\begin{abstract}
  We propose a continuous wavelet transform based non-parametric
  foreground subtraction method for the detection of redshifted 21 cm
  signal from the epoch of reionization. This method works based on
  the assumption that the foreground spectra are smooth in frequency
  domain, while the 21 cm signal spectrum is full of saw-tooth-like
  structures, thus their characteristic scales are significantly
  different. We can distinguish them in the wavelet coefficient space
  easily and perform the foreground subtraction. Compared with the
  traditional spectral fitting based method, our method is more
  tolerant to complex foregrounds. Furthermore, we also find that when
  the instrument has uncorrected response error, our method can also
  work significantly better than the spectral fitting based
  method. Our method can obtain similar results with the Wp smoothing
  method, which is also a non-parametric method, but our method
  consumes much less computing time.
\end{abstract}

\section{Introduction}
In the end of the dark ages, with the formation of the first
generation of stars and/or quasars, the neutral hydrogen in the
universe began to reionize. The redshifted 21 cm signal is one of the
most important signatures for detecting the epoch of reionization
(EoR), which has been a new frontier of astrophysics in recent
years. The 21 cm line emission/absorption from the EoR has been
redshifted to meter wave band. Several facilities that cover the meter
wave band have been or will be built to explore EoR through the
redshifted 21 cm signal, such as the 21 Centimeter Array
(21CMA\footnote{http://21cma.bao.ac.cn}), the Low Frequency Array
(LOFAR\footnote{http://www.lofar.org}), the Murchison Wide-field Array
\footnote{http://www.mwatelescope.org}, and the Square Kilometre Array
(SKA\footnote{http://www.skatelescope.org}). The predicted brightness
temperature of the EoR 21 cm signal is a few $10^{-2}$ K
\citep[e.g.,][]{2008ApJ...676....1B, 2008PhRvD..78j3511P,
  2004MNRAS.347..187F}, while that of the foreground emission from the
Milky Way and extragalactic radio sources reaches $10^2$ K and even
higher \citep[e.g.,][]{2008MNRAS.391..383G, 2006ApJ...650..529W},
i.e., brighter than the EoR 21 cm signal by four orders of
magnitude. In order to detect the EoR 21 cm signal, the biggest
challenge that one must overcome is how to subtract the strong
foreground emission.

A dozen of foreground subtraction methods have been proposed. A
typical one is the polynomial fitting method \citep[e.g.,
][]{2006ApJ...650..529W}, the principle of which is rather
straightforward. The spectrum of the total signal is fitted with a low
order (e.g., second or third) polynomial in logarithmic space, and the
residual is regarded to be the sum of the 21 cm signal and the
instrumental noise. This kind of method can be applied either in $uv$
space or real space. Some derivative methods have been released, for
example, \cite{2013ApJ...763...90W}. Another important sort of
foreground subtraction method is the non-parametric method
\citep[e.g.,][] {2009MNRAS.397.1138H,2013MNRAS.429..165C}. Unlike the
fitting based method, the non-parametric method does not assume the
detailed form (usually a polynomial-like form in logarithmic space) of
the foreground spectrum.

Both kinds of methods assume that the foreground spectrum is smooth in
frequency space, or more generally, the emissions of different
frequencies are strongly correlated, while the EoR 21 cm signal is
full of saw-tooth-like structures. In other words, the characteristic
scales of the foreground and EoR signals are different in their radio
spectrum space. The power spectral density of the foreground signal is
mainly contributed by large scale components in radio frequency space,
while that of the EoR signal is mainly contributed by small scale
components \citep[e.g., $\lesssim1$ MHz at $z=8$, corresponding to
$\lesssim8$ comoving Mpc;][]{2007ApJ...669..663M}. This difference can
be used to distinguish between them. For general one-dimensional signals
like time series and spectrum, mature mathematical tools have been
invented, and the wavelet transform is one among them.

In this work , we study the feasibility of using continuous wavelet
transform \citep[CWT, e.g., ][]{1992tlw..conf.....D} to subtract the
strong foreground emission in radio spectrum space. We describe the
simulation of the foreground, 21 cm, and thermal noise signals in
\S\ref{sec:simu}, give a brief introduction to the CWT that we use in
\S\ref{sec:wt}, test the foreground subtraction with simulated signals
in \S\ref{sec:sub}, discuss our results in \S\ref{sec:discussion}, and
conclude our work in \S\ref{sec:conclusion}. Throughout the paper, we
adopt $H_0=100h$ km s$^{-1}$ Mpc$^{-1}$, where $h=0.71$, $\Omega_{\rm
  M}=0.27$, $\Omega_\Lambda=0.73$, and $\Omega_b=0.044$ \citep[e.g.,
][]{2003ApJS..148..175S}.

\section{Simulation of the Low-frequency Radio Spectra}
\label{sec:simu}
\subsection{Redshifted 21 cm Signal from the Epoch of Reionization}
\label{ssec:21cm}
According to \cite{2006ApJ...650..529W} and references therein, we
simulate the redshifted 21 cm signal from the EoR
based on the theoretical three-dimensional power spectrum, which is
represented as
\begin{gather}
  P_{\rm 3D, 21 cm}(k,z)=(16{\rm~mK})^2\frac{1}{h^2}\left (\frac{\Omega_{\rm b}h^2}{0.02}\right)^2\frac{1+z}{10}\frac{0.3}{\Omega_{\rm M}}\notag\\
  \times \left \{[1-x_{\rm e}^2(z)]^2+b^2(z)e^{-k^2R^2(z)}x_{\rm
      e}^2(z)\right\}P_{\rm 3D, matter}(k,z),
\end{gather}
where $P_{\rm 3D, matter}$ is the three-dimensional matter power spectrum
at redshift $z$, $x_{\rm e}(z)$ is the average ionization fraction,
$b(z)$ is the mean halo bias, and $R(z)=100[1-x_{\rm e}(z)]^{-1/3}$
kpc is the mean radius of the ionized patches in H{\sc II} regions. We
use the $b(z)$ and $x_{\rm e}(z)$ values calculated from the fiducial
reionization model of
\cite{2003ApJ...598..756S,2005ApJ...625..575S}. In this work, we
consider a redshift interval $\Delta z\ll 1$, so that the power
spectrum can be regarded as uniform. In the narrow redshift range, we
are able to calculate the one-dimensional power spectrum by the
formula\footnote{The $n$-dimensional Fourier transform convention that
  we use here is
  \begin{gather}
    F_{k}(\mathbf{k})=\left(\frac{1}{L}\right)^n\int F(\mathbf{x})\exp(i\mathbf{k}\cdot\mathbf{x})d^nx\notag\\
    F(\mathbf{x})=\left(\frac{2\pi}{L}\right)^n\int F_{k}(\mathbf{k})\exp(-i\mathbf{k}\cdot\mathbf{x})d^nk\notag
  \end{gather}} 
\begin{gather}
  P_{\rm 1D, 21 cm}(k,z)=\frac{1}{2\pi}\int_{k}^{\infty}P_{\rm 3D, 21 cm}(k^\prime,z)k^\prime d k^\prime,
\end{gather}

according to \cite{1999coph.book.....P}. The line-of-sight
distribution of the redshifted 21 cm line brightness temperature,
i.e., the radio spectrum in an interval $(\nu_0-\Delta\nu/2,\nu_0+\Delta\nu/2)$
(corresponding to a redshift interval $(z_0-\Delta z/2,z_0+\Delta z/2)$, where
$z=1420.4$ MHz/$\nu-1$) can be represented as the sum of a series of
Fourier bases as
\begin{gather}
  T_{\rm 21 cm}(x_n)=\sum_{q=0}^{N-1}\left[ A_q\cos\left(\frac{2\pi
        q}{L}x_n\right)+B_q\sin\left(\frac{2\pi
        q}{L}x_n\right)\right],
\end{gather}
where $L$ is the comoving space scale corresponding to $\Delta z$,
$A_q$ and $B_q$ are random variables that independently follow an
identical normal distribution $\mathcal{N}(0,\sqrt{2P_{\rm 1D, 21
    cm}(k,z_0)/L})$, $k=2\pi q/L$ is the wave number, $x_n$ ($n=1,
2,\cdots,N$) is the line-of-sight location related to frequency
$\nu_n$ under linear approximation, and $N$ is the number of frequency
channels involved. $N$ is chosen to be large enough so that most power
of the power spectrum $P_{\rm 1D, 21 cm}(k,z)$ is enclosed within the
wave number interval $2\pi/L\leq k\leq 2\pi N/L$. In this work, we
take $z_0=8$ as an example, and let $\Delta\nu=20$ MHz, $N=500$, so
that the frequency resolution is $d\nu=\Delta\nu/N=40$ kHz, which is a
realistic value that can be achieved by most existing and upcoming EoR
probing facilities.

\subsection{Foreground Emission}
\label{ssec:fg}
\begin{figure}
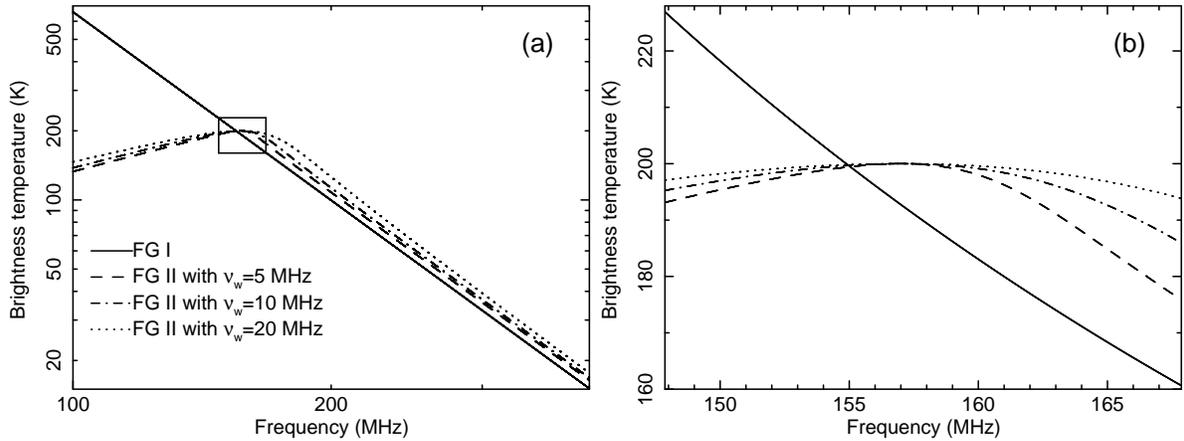

  \begin{center}
    \includegraphics[width=.35\textwidth,angle=270]{fg_large.ps}
    \includegraphics[width=.35\textwidth,angle=270]{fg_small.ps}
  \end{center}
  \caption{\label{fig:fg} Spectra of the FG I and FG IIs with
    different conjunction bandwidth $\nu_w$. The solid box in (a) is
    enlarged to be shown in (b).}
\end{figure}

Compared with relatively high frequency radio observations ($\ge 1.4$
GHz), low frequency radio observations are far from mature. Detailed
properties of radio emission from radio sources are still not
clear. It is not appropriate to place too strong assumptions over the
foreground emission. Nevertheless, the smoothness of the foreground
radio spectra is widely accepted \citep[e.g., ][]{1999A&A...345..380S,
  2011MNRAS.413.2103P}. We also accept this assumption in present
work. To study the effect of our proposed method, we decide to test
two kinds of foregrounds (Fig. \ref{fig:fg}). One is based on high
frequency and relatively limited low frequency observations of radio
sources, which has a set of analytic formulas deduced from clear
physical mechanisms. Another foreground model that we use is composed
purely mathematically, which is used to test the tolerance of the
foreground subtraction method. The two kinds of foregrounds are
described in detail as follows.

For the first kind of foreground, we simulate spectra following our
previous work \citep[][and references therein]{2010ApJ...723..620W},
which includes the radio emission from the Milky Way, galaxy clusters,
and discrete extragalactic radio sources (i.e., star-forming galaxies,
radio-quiet AGNs, and radio-loud AGNs). We call it ``FG I'' in the
following sections.

For the second kind of foreground, we assume the spectrum to possess
two different power law indices ($\alpha_1=1$ and $\alpha_2=-2.7$) in
lower and higher frequency bands, respectively. The spectrum turnover
feature in this foreground model originates from the absorption in
many sources \citep[][]{2008MNRAS.390L..43G, 2004A&A...424...91E,
  2003A&A...402..171T, 2000A&A...363..887D,
  1999MNRAS.305..492D,2012ApJ...760...77A}. Several absorption
mechanisms have been proposed, and their detailed spectra are
different. Nevertheless, we just compose a pure mathematical model to
represent the general feature when absorption happens.  Note that in
the simulation of FG I, absorption mechanisms have been considered for
different kinds of sources; we are only making the feature of
absorption more obvious to make the foreground more complex for the
test purpose. The slope changes smoothly around the turnover frequency
$\nu_t=157$ MHz ($\nu_t$ is randomly chosen here only for the test
purpose), and the sharpness of the turnover point is described by a
width parameter $\nu_w$. The spectrum is described by following
equations as
\begin{gather}
  T_{\rm b}(\nu)=T_{\rm b}(\nu_t)\left
    (\frac{\nu}{\nu_t}\right)^{(1-w)\alpha_1+w\alpha_2}\label{equ:fg2}
\end{gather}
where 
\begin{gather}
  w=\frac{2\arctan[(\nu-k\nu_t)/\nu_w]/\pi+1}{2},\\
  k=1-\frac{\nu_w}{\nu_t}\tan\frac{\pi(\alpha_1+\alpha_2)}{2(\alpha_1-\alpha_2)},
\end{gather}
and $T_{\rm b}(\nu_t)$ is the brightness temperature at the turnover
frequency. The above equation ensures that the spectrum reaches a peak
at the turnover frequency. We call it ``FG II'' in the following sections.

\subsection{Instrumental Noise}
\label{ssec:noise}
Instrumental noise is crucial for the detection of the EoR 21 cm
signal. When the instrumental noise is significantly lower than the 21
cm signal, the detailed form of 21 cm signal can be derived, and only
statistical information can be obtained otherwise. We will test the
method with two different noise levels, one is higher than the 21 cm
signal, and another one is lower. We use 60 mK as the higher noise
level like we have used in our previous work
\citep{2013ApJ...763...90W}, which is based on the parameters of the
21CMA. Other working facilities such as LOFAR can also reach this
noise level. For example, according to the parameters used in
\cite{2010MNRAS.405.2492H} and \cite{2006ApJ...650..529W}, when the
core area of LOFAR is configured with a channel bandwidth of several
$10^1$ kHz, and the observing time reaches month level, its noise
level will reach several $10^1$ mK. The lower noise level is taken to
be $6$ mK, which is calculated according to the future SKA core region
parameters as follows.

According to \cite{2001isra.book.....T}, the brightness temperature
measurement error of an interferometer is calculated as
\begin{gather}
  \Delta T_{\rm b}=\frac{\lambda^2 T_{\rm sys}}{A_{\rm e}}\frac{1}{\Omega_{\rm beam}\sqrt{d\nu\tau n(n-1)}}\notag\\
  \approx\frac{\lambda^2T_{\rm sys}}{nA_{\rm e}}\frac{1}{\Omega_{\rm beam}\sqrt{d\nu\tau}},\label{equ:noise}
\end{gather}
where $\lambda$ is the wavelength, $T_{\rm sys}$ is the system
temperature, $A_{\rm e}$ is the effective area of one antenna,
$\Omega_{\rm beam}$ is the solid angle of the synthesized beam, $d\nu$
is the channel bandwidth, $\tau$ is the observing time, and $n$ is the
number of antennas. According to the design of SKA, there will be
$\eta=30\%$ of the total collecting area ($n_{\rm total}A_{\rm e}$)
within the central 1--2 km. Here, we assume the baseline length to be
$L=1.5$ km, so that $\Omega_{\rm
  beam}\approx\frac{\pi}{4}\frac{\lambda^2}{L^2}=1.40\times10^{-6}$
sr. We rewrite Equation (\ref{equ:noise}) as
\begin{gather}
  \Delta T_{\rm b}\approx\frac{\lambda^2T_{\rm sys}}{\eta n_{\rm
      total}A_{\rm e}}\frac{1}{\Omega_{\rm beam}\sqrt{d\nu\tau}}.
\end{gather}
For the SKA, the parameter $(n_{\rm total}A_{\rm e})/T_{\rm sys}=5000$
m$^2$ K$^{-1}$. Finally, we have
\begin{gather}
  \Delta T_{\rm b}\approx\frac{4L^2}{\pi}\frac{T_{\rm sys}}{n_{\rm total}A_{\rm e}}\frac{1}{\eta\sqrt{d\nu\tau}}\\
  =6~{\rm mK}\left (\frac{40~{\rm
        kHz}}{d\nu}\right)^{1/2}\left(\frac{30~{\rm
        days}}{\tau}\right)^{1/2}.
\end{gather}
Given a channel bandwidth of $40$ kHz, and a total observing time of 1
month, the brightness temperature noise level can reach a few mK. In
following analysis, we use $6$ mK as the lower noise level.

\section{Continuous Wavelet Transform}
\label{sec:wt}
We make a brief introduction in the context of our problems,
according to the detailed review by \cite{1992tlw..conf.....D}. The
spectra from different sorts of sources (i.e., continuous spectra
mainly from synchrotron emission and line emission/absorption from the
neutral hydrogen in different epochs of the universe) possess
different characters. Continuous spectra are smooth in radio frequency
space, while line emission/absorption spectra are full of
saw-tooth-like structures. One can quantify the above difference and
separate them in frequency space.

\subsection{From Short-time Fourier Transform to Continuous Wavelet
  Transform}
For stationary signal, one possible method to quantify the above
difference is Fourier transform. Continuous spectra are composed of
more low frequency components, while saw-tooth-like line
emission/absorption spectra are composed of more high frequency
components. So in principle, the smooth spectra and the saw-tooth-like
spectra can be separated with a pair of low-pass and high-pass
filters. However, the assumption that the spectra are stationary
signals is not safe enough.

To handle non-stationary signal, the real space resolution should be
kept. The short-time Fourier transform (STFT) is a modification of
traditional Fourier transform, which partly keeps the real space
resolution. The STFT of a real space signal $h(t)$ can be defined as
\begin{gather}
  (\mathcal{F}h)(\tau,\omega)=\int_{-\infty}^{\infty}h(t)w(t-\tau)e^{-i\omega
    t}dt,\label{equ:stft}
\end{gather}
where $w(t)$ is the window function, which should meet the
normalization requirement
\begin{gather}
  \int_{-\infty}^{\infty}w(\tau)d\tau=1.
\end{gather}
As a commonly used window function, the Gaussian window function is
usually defined as
\begin{gather}
  w_{g}(x)=\frac{1}{\sqrt{2\pi}s}e^{-\frac{x^2}{2s^2}},\label{equ:gaussian_window}
\end{gather}
where the parameter $s$ determines the real space resolution. By using
the Gaussian window, the STFT becomes
\begin{gather}
  (\mathcal{F}h)(\tau,\omega)=\int_{-\infty}^{\infty}h(t)
  \frac{1}{\sqrt{2\pi}s}e^{-\frac{(t-\tau)^2}{2s^2}} e^{-i\omega
    t}dt.\label{equ:gaussian_stft}
\end{gather}
Although the STFT partly keeps the real space resolution, it is not
adaptively determined by the frequency $\omega$, i.e., a global and
fixed parameter $s$, which represents that the width of the window
function is used for all frequencies.

The CWT can overcome the above difficulty. According to e.g.,
\cite{1998BAMS...79...61T}, the one-dimensional CWT is defined as
\begin{gather}
  W_{x,\psi}(\tau,s)=\int_{-\infty}^{\infty}h(t)\frac{1}{\sqrt{|s|}}\psi^*(\frac{t-\tau}{s})dt,\label{equ:cwt}
\end{gather}
where $h(t)$ is the real space signal to be transformed, $\psi$ is
called the mother wavelet function, $\tau$ and $s$ represent the real
space and scale indices of the wavelet coefficient
$W_{x,\psi}(\tau,s)$, respectively. The mother wavelet function
$\psi\in L^2(\mathbb{R})$ (quadratically integrable function) should
meet the requirement
\begin{gather}
  0<C_{\psi}\equiv\int_{-\infty}^{\infty}\frac{|\Psi(\omega)|}{|\omega|}d\omega<\infty,
\end{gather}
where $\Psi$ is the Fourier transform of $\psi$, and $C_{\psi}$ is
called the admissibility constant. According to Equation
(\ref{equ:cwt}), given a certain scale $s$, the wavelet transform is
actually the cross-correlation between $\psi_{s}(t)=\psi(t/s)/\sqrt{|s|}$ and the
real space signal $h(t)$, so that according to the cross-correlation
theorem, it can be calculated efficiently in Fourier space as
\begin{gather}
  W_{x,\psi}(\tau,s)=\mathcal{F}^{-1}\{\Psi_s^*\cdot H\},\label{equ:cwt_fft}
\end{gather}
where $\Psi_s$ and $H$ are the functions $\psi_s$ and $h$ in Fourier
space. In practice, the Fourier transform can be calculated with any
fast Fourier transform package discretely. In this work, we implement
the transform with the {\it FFTW3} package \citep{FFTW05}.

The most commonly-used mother wavelet functions include Morlet
\citep{1982Geop...47..203M}, Paul \citep[e.g.,
][]{1998BAMS...79...61T}, DOG \citep[e.g., ][]{1998BAMS...79...61T},
etc. We prefer the Morlet mother wavelet function, which is defined as
\begin{gather}
  \psi(x)=\pi^{-1/4}e^{2\pi if_0 x}e^{-x^2/2},
\end{gather}
where $f_0$ is the frequency parameter, for the following reasons.  First,
according to the comparison among a variety of wavelet functions in
\cite{1998BAMS...79...61T} Morlet mother wavelet function has the
advantage that it can obtain better frequency resolution (not to be
confused with the radio frequency, here the term ``frequency''
represents the scale of the signal component), but poorer real space
resolution (in our context, it is radio frequency resolution). In this
work, we are more interested in the separation between smooth
continuum emission and saw-tooth-like line emission according to their
different characteristic scales; in other words, the frequency
resolution (i.e., the scale resolution) is relatively more important
to us.  Second, the form of the Morlet mother wavelet function can be
obviously regarded as the product of a Gaussian window function and
the Fourier base, so that the Morlet wavelet transform can be treated
as an STFT that has a frequency-dependent window function. In other
words, the Morlet wavelet transform can be smoothly introduced by
slightly modifying the Gaussian window STFT.

\begin{figure}
  \begin{center}
    \includegraphics[height=.75\textwidth,angle=270]{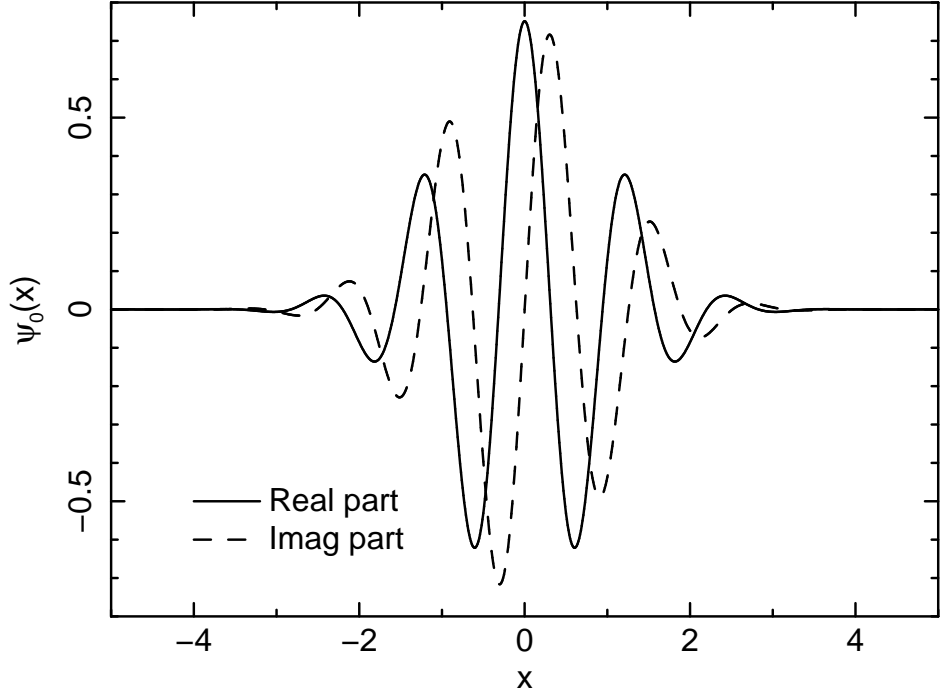}
  \end{center}
  \caption{\label{fig:wavelet} Morlet mother wavelet function with
    $f_0=1$ channel$^{-1}$.}
\end{figure}

The $f_0$ parameter is chosen to be 1 channel$^{-1}$ (i.e., $25$
MHz$^{-1}$) in this work to match the radio frequency resolution. We
plot the Morlet mother wavelet function that we use here in Figure
\ref{fig:wavelet}.  According to \cite{1992tlw..conf.....D}, the
inverse CWT is defined as
\begin{gather}
  h(t)=\frac{2}{C_{\psi}}\int_0^\infty\left
    [\int_{-\infty}^{\infty}W_{x,\psi}(\tau,s)\frac{1}{\sqrt{|s|}}\psi(\frac{t-\tau}{s})d\tau\right
  ]\frac{ds}{s^2}.\label{equ:icwt}
\end{gather}
Given a certain scale $s$ the inner part of the double
  integration is actually a convolution between $W_{x,\psi}$ and
  $\psi_s$, so that can also be calculated efficiently in Fourier
  space just like what we have done in Equation (\ref{equ:cwt_fft}),
  as
\begin{gather}
  h(t)=\frac{2}{C_{\psi}}\int_0^{\infty}\mathcal{F}^{-1}\{\mathcal{F}\{W_{x,\psi}\}\cdot\Psi_{s}\}(t,s)\frac{ds}{s^2}.
\end{gather}
After the signal is filtered in the wavelet coefficient space, we will
use this equation to transform it back to real space.

\subsection{Boundary Effects}
\label{ssec:boundary}
According to the definition of CWT (Equation (\ref{equ:cwt})), the
input signal is assumed to be infinite in real space, which however
does not hold in practice. In other words, when trying to use CWT to
subtract the foreground in radio frequency space, the radio bandwidth
is not infinitely broad. There are several methods to extend a finite
signal to an infinite one to meet the definition of CWT.  Filling
zeros, period extension, and symmetric extension are the most common
ones. We have tested all the three extension methods above and find no
significant difference among them. We provide a further discussion
about handling the boundary effect in \S\ref{ssec:extension}. Although
general extension methods will introduce discontinuity, and may
contaminate the transformed signal, it will not significantly affect
the final results for the following reason. Almost all working and
upcoming facilities that aim to detect the 21 cm signals from the EoR
have a much broader bandwidth than that is required in most
conditions. For example, when calculating the one-dimensional H{\sc I}
power spectrum of a certain redshift from the radio spectra, the
adopted bandwidth is usually limited to several MHz to ensure the
uniformity of the power spectrum, we can perform the subtraction over
a larger bandwidth than needed, and only use the bandwidth section
that is less affected by the boundary effects. So, we simply use the
period extension method to handle the finite signal bandwidth.

\section{Subtraction of Foreground Signal}
\label{sec:sub}
The difference between the distribution of significant wavelet
coefficients of the foreground and 21 cm signals can be used to
distinguish between them. In the following sections, we first study the
characters of the wavelet coefficients of the foreground and 21 cm
signals, respectively. After that, we test the wavelet based method of
subtracting a strong foreground.

\subsection{Wavelet Coefficients of Different Kinds of Sources}
\label{ssec:coeff_diff}
We first study the characters of the wavelet coefficients of the
foreground that been simulated above (\S\ref{ssec:fg}). We show the
absolute value of the wavelet coefficients of the FG I and FG IIs with
$\nu_w=$5, 10, and 20 MHz in Figure \ref{fig:wt_fg}.\footnote{Note
  that for presentation purpose all the wavelet coefficients shown in
  figures are multiplied by $10^3$.} The most impressive character of
the coefficients of the four foregrounds is that the most significant
coefficients are contributed by the boundary effect of the data. In
our tests, it is hard to disentangle the boundary effect and the
contribution from the foreground signal itself, because both of them
are more prominent on large scales. However this is not a serious
problem, since what we actually want to obtain is not the foreground
signal itself.

\begin{figure}
  \begin{center}
    \includegraphics[height=.75\textwidth,angle=270]{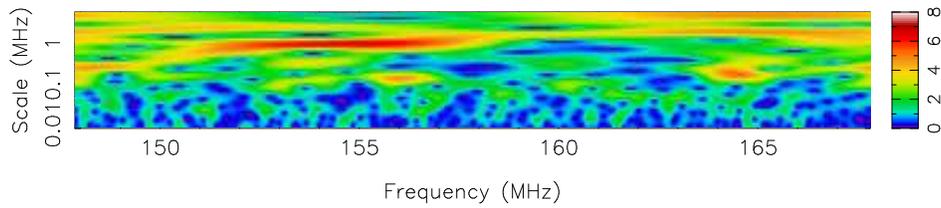}
  \end{center}
  \caption{\label{fig:wt_21} The absolute values of the wavelet
    coefficients of one realization of the simulated EoR 21 cm
    signal.}
\end{figure}

Then we study the behavior of the 21 cm signal (\S\ref{ssec:21cm})
with the CWT. We randomly choose one realization of the simulated 21
cm signal, and calculate the wavelet coefficients. The result is shown
in Figure \ref{fig:wt_21}. Different from those of the foregrounds,
the coefficients of the 21 cm signal appear to be much more prominent
in small-scale regions, and much less affected by the boundary
effects.

\subsection{Filtering Out the Foreground Signal}
\label{ssec:filtering}
\begin{figure}
  \begin{center}
    \includegraphics[height=.75\textwidth,angle=270]{coi.ps}\\
    \includegraphics[height=.75\textwidth,angle=270]{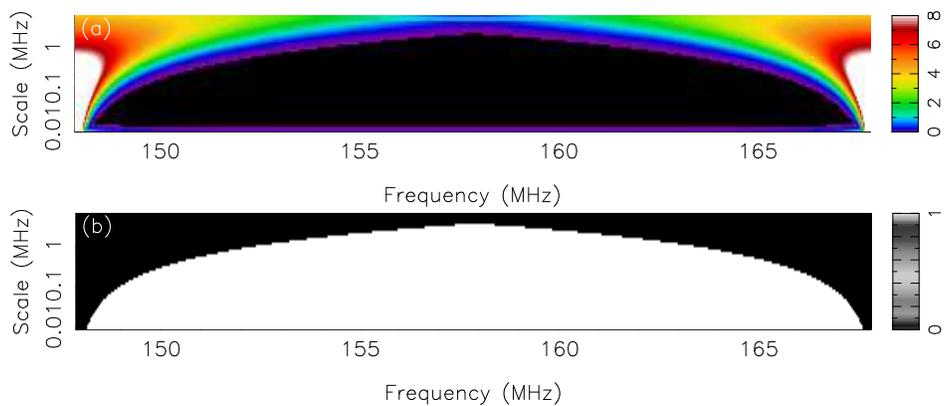}
  \end{center}
  \caption{\label{fig:coi} (a) The absolute values of the wavelet
    coefficients of Equation (\ref{equ:dd}), which are used to
    determine regions contaminated by the boundary effect. (b) The
    mask that is used to filter out the wavelet coefficients.}
\end{figure}

As we have noted that the distribution of the significant coefficients
of the foreground signal and the 21 cm signals are different
(\S\ref{ssec:coeff_diff}), we can utilize this character to filter out
the foreground signals. Because the significant coefficients of the
smooth foregrounds are mainly contributed by the discontinuity of the
data boundary, the simplest way is to exclude the regions affected by
the boundary effect. To determine the regions to be excluded, we
calculate the wavelet coefficients of function
\begin{gather}
  T_{d}(\nu)=\delta_c(\nu-\nu_{\min}),\label{equ:dd}
\end{gather}
as shown in Figure \ref{fig:coi}a, where $\delta_c$ is the Dirac delta
function and $\nu_{\min}=147.8$ MHz is the lower limits of our test
band. Strictly speaking, the step function is more suitable for
representing the discontinuity near the boundary; however, in our
practical condition, we choose the Dirac delta function due to its
localization property. In detail, because we use the period signal
extension method, it is impossible to compose such a step function
with the jumps at $\nu_{\rm min}$ and $\nu_{\rm max}$ only. On the
other hand, the Dirac delta function can be regarded as the derivative
of step function and according to the property of Fourier transform,
the difference between the Morlet wavelet transforms of Dirac delta
and step functions is only a slowly varying factor, which only has a
minor impact on our results.  Note that since we use a period signal
extension method, the discontinuity of the upper limit of the signal
will also be reflected by Equation (\ref{equ:dd}). For any given scale
$s$, the absolute value of the wavelet coefficient peaks at
$\nu_{\min}$ and $\nu_{\max}$, i.e., the boundary of the data
series. The wavelet coefficients of the $T_d(\nu)$ can be used to
recognize the region that is significantly affected by the boundary
effect.  We can empirically define a threshold for each scale to be
$10^{-2}$ of the peak value. The regions, where the absolute value of
the wavelet coefficient is above the threshold should be marked to be
excluded. Because the absolute value of the coefficient decreases
exponentially as the distance from the boundary increases, the masked
region is relatively insensitive to the value of the threshold.  With
this standard, we generate the mask for filtering out the foreground
(Fig. \ref{fig:coi}b).

\subsection{Results}
\label{ssec:results}
\begin{figure}
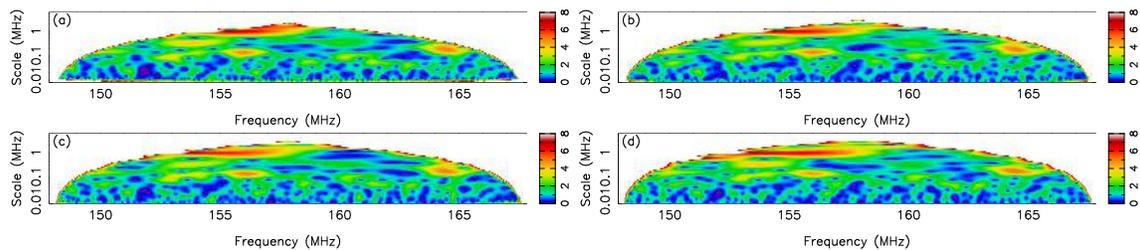

  \begin{center}
    \includegraphics[height=.45\textwidth,angle=270]{wt_total1.ps}
    \includegraphics[height=.45\textwidth,angle=270]{wt_total2.ps}\\
    \includegraphics[height=.45\textwidth,angle=270]{wt_total3.ps}
    \includegraphics[height=.45\textwidth,angle=270]{wt_total4.ps}
  \end{center}
  \caption{\label{fig:wt_total} (a)-(d) The absolute values of the
    wavelet coefficients of the total signal of FG I and FG IIs with
    $\nu_w=$5, 10, and 20 MHz, respectively.}
\end{figure}

\begin{figure}
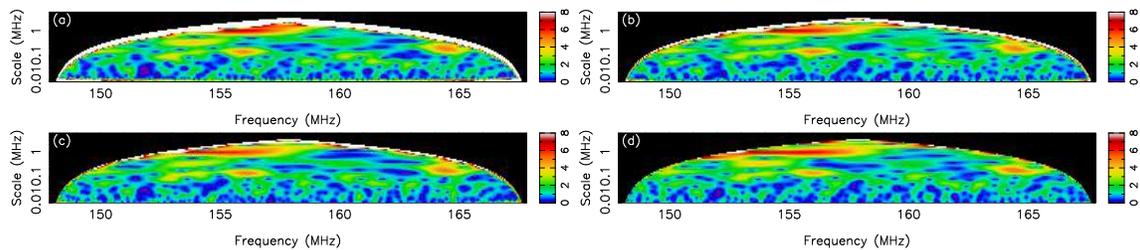

  \begin{center}
    \includegraphics[height=.45\textwidth,angle=270]{wt_filtered1.ps}
    \includegraphics[height=.45\textwidth,angle=270]{wt_filtered2.ps}\\
    \includegraphics[height=.45\textwidth,angle=270]{wt_filtered3.ps}
    \includegraphics[height=.45\textwidth,angle=270]{wt_filtered4.ps}
  \end{center}
  \caption{\label{fig:wt_filtered} (a)-(d) The absolute values of
    the wavelet coefficients of total signals with FG I and FG IIs
    with $\nu_w=$5, 10, 20 MHz, respectively, filtered by using the
    mask shown in Figure \ref{fig:coi}b.}
\end{figure}

Multiplying the wavelet coefficients of the total signal
(Fig. \ref{fig:wt_total}) by the mask (Fig. \ref{fig:coi}b), we derive
the filtered coefficients as is shown in Figure
\ref{fig:wt_filtered}. Then by using Equation (\ref{equ:icwt}), we
reconstruct the filtered 21 cm signal. To further avoid the boundary
effect, we exclude the signal with $\nu<\nu_{\rm min}+5$ MHz and
$\nu>\nu_{\rm max}-5$ MHz. Note that the bandwidth that is cut here is
chosen empirically, considering the trade-off between the available
bandwidth and the boundary effect. Although it seems that we have
wasted half of the total band, actually in real observations we can
move the subtraction band, i.e., $(\nu_{\rm min},\nu_{\rm max})$
continuously in the range of the total instrument band, so that most
of the frequency range can be used. Our result shows that a total
observation bandwidth of $20$ MHz enables us to detect the HI 21 cm
line emission distribution in one redshift period. Broader bandwidth
should enable us to study wider redshift range.

\begin{figure}
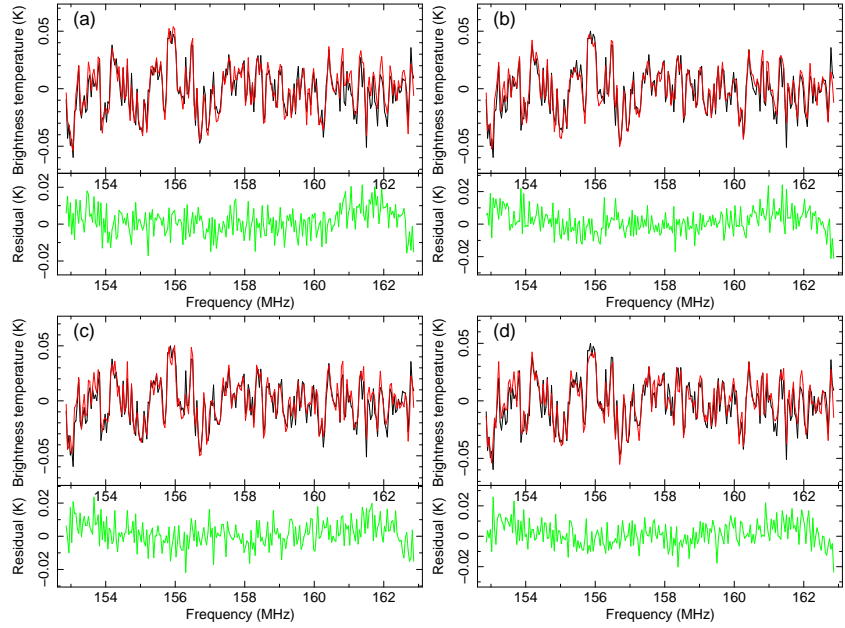

  \begin{center}
    \includegraphics[height=.33\textwidth,angle=270]{result1.ps}
    \includegraphics[height=.33\textwidth,angle=270]{result2.ps}\\
    \includegraphics[height=.33\textwidth,angle=270]{result3.ps}
    \includegraphics[height=.33\textwidth,angle=270]{result4.ps}
  \end{center}
  \caption{\label{fig:result} Foreground subtraction results of the
    wavelet based method with $\Delta T_{\rm b}=6$ mK. (a)-(d) The
    results of using FG I and FG IIs with $\nu_w=$5, 10, and 20 MHz,
    respectively. The black, red, and green lines show the input 21 cm
    signal, the output 21 cm signal, and the residual, respectively.}
\end{figure}

Given the noise level calculated for the SKA core region (i.e.,
$\Delta T_{\rm b}=6$ mK in \S\ref{ssec:noise}), we test our method on
the simulated foregrounds and 21 cm signals. Suppose that the noise
level is significantly lower than the EoR 21 cm signal, we can obtain
the spectrum or the distribution of H{\sc I} along the line-of-sight
of a certain sky region covered by a single beam. Typical
reconstructed results with the above four foregrounds are shown in Figure
\ref{fig:result}.

\begin{figure}
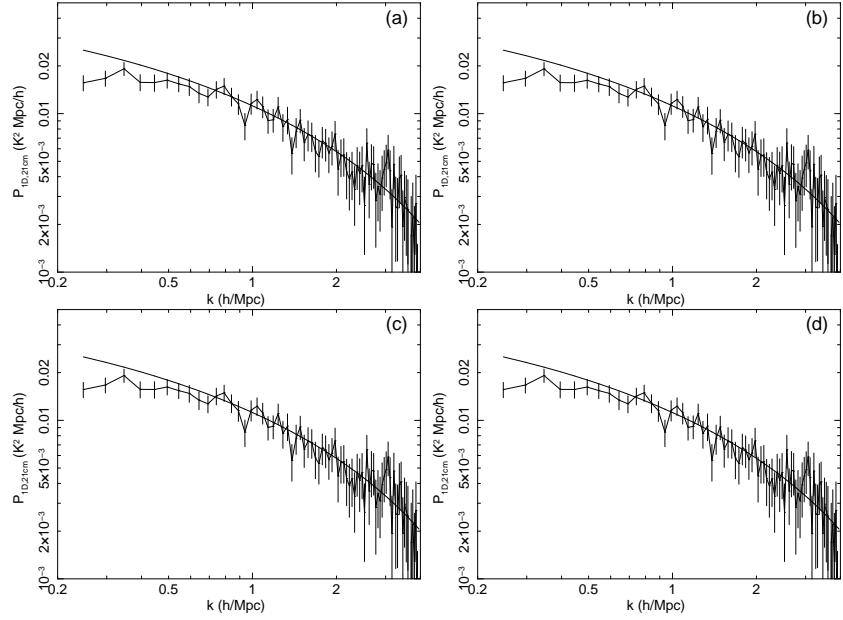

  \begin{center}
    \includegraphics[height=.33\textwidth,angle=270]{ps1.ps}
    \includegraphics[height=.33\textwidth,angle=270]{ps2.ps}\\
    \includegraphics[height=.33\textwidth,angle=270]{ps3.ps}
    \includegraphics[height=.33\textwidth,angle=270]{ps4.ps}
  \end{center}
  \caption{\label{fig:ps} Reconstructed power spectra of the EoR
    21 cm signals by using the wavelet based method, with $\Delta
    T_{\rm b}=60$ mK. (a)-(d) The results of using FG I and FG IIs
    with $\nu_w=$5, 10, and 20 MHz, respectively. The solids line and
    the data points are the theoretical and estimated power spectra,
    respectively.}
\end{figure}

If the brightness temperature noise is comparable with the EoR 21 cm
signal, we can only obtain its statistical properties such as the
power spectrum. It is obvious that the information that we can extract
from one-dimensional power spectrum is relatively limited when
compared with three-dimensional power spectrum, but we still test the
one-dimensional power spectrum here. There are two reasons. The first
reason is that in this work, we only simulate the 21 cm spectrum on
each pixel, rather than a three-dimensional data cube, so that with
our simulated data, we cannot calculate the three-dimensional power
spectrum. Future simulations by using codes such as 21CMFAST
\citep{2011MNRAS.411..955M} may enable us to perform more complete
tests, which will be a part of our future work. The second reason is
that the calculation of one-dimensional power spectrum is relatively
less dependent on instrument parameters and is relatively simple. To
calculate three-dimensional power spectrum, one must compose data cube
in real space and transform it into Fourier space, during which the
survey strategy, especially the shape and area of the sky coverage
must be considered, while one-dimensional power spectrum only requires
measuring the radio spectrum on each interesting pixel, and does not
need to consider how these pixels are distributed, so that the results
should be more general.  Assuming a noise level of $60$ mK
(\S\ref{ssec:noise}), we calculate the one-dimensional 21 cm power
spectrum by averaging the line-of-sight power spectra from 1000 beams,
and subtract the predicted instrumental noise power spectrum. We show
the results in Figure \ref{fig:ps}. We find that there is some power
leakage, which is especially severe in the small wave number end
($k<0.4h$ Mpc$^{-1}$). This is mainly caused by the filtering
strategy, and can be corrected as described in the following.

\begin{figure}
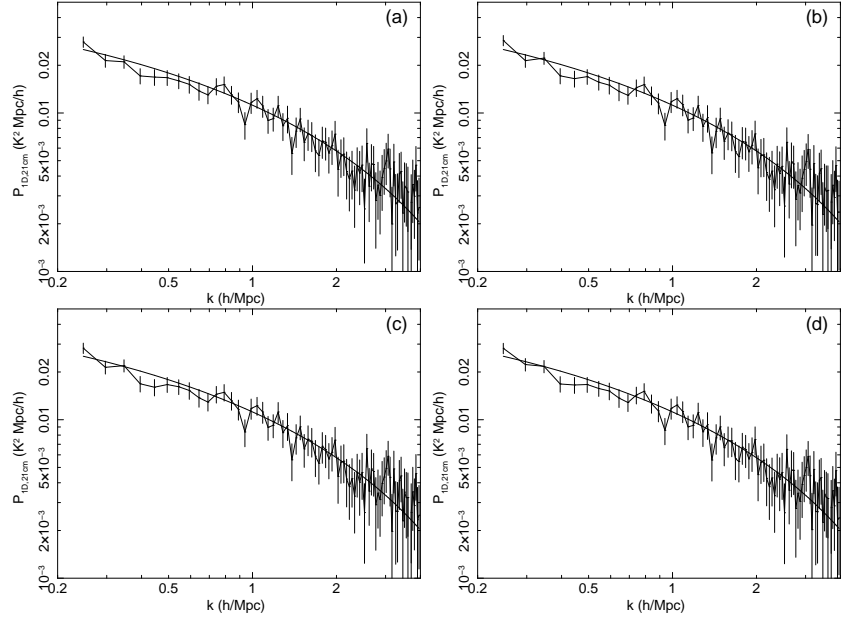

  \begin{center}
    \includegraphics[height=.33\textwidth,angle=270]{ps1_corr.ps}
    \includegraphics[height=.33\textwidth,angle=270]{ps2_corr.ps}\\
    \includegraphics[height=.33\textwidth,angle=270]{ps3_corr.ps}
    \includegraphics[height=.33\textwidth,angle=270]{ps4_corr.ps}
  \end{center}
  \caption{\label{fig:ps_corr} Same as Figure \ref{fig:ps}, but
    corrected for the power leakage in the small wave number end. }
\end{figure}

There are at least two methods to correct the power leakage, which is
especially severe at the small wave number end. Both of the two
methods work by multiplying a correction factor with the produced
power spectrum, which is the function of wave number $k$. In the first
method, we can feed a standard signal, the power spectrum of which is
known in advance, into the foreground subtraction program, and
calculate the power spectrum of the output signal, which is then
compared with that of the input signal, and calculate the correction
factor. This method can be named as the closed loop method. In the
second method, the correction factor is calculated as the ratio of the
total bandwidth of the input signal to that of the bandwidth after
masked for a certain scale, i.e., the ratio of the total bandwidth to
the width of the white region in Figure \ref{fig:coi}b at different
scales. Then according to \cite{Kirby2005846}, the wavelet scale $s$
of Morlet wavelet transform has a Fourier wave number counterpart
$2\pi f_0/s$. So that the correction factor can be converted to a
function of the wave number and can be applied to correct the power
spectrum. This method can be named as the open loop method. In
principle, the closed loop method should be more precise since it
avoids the issue of converting the wavelet scale $s$ to the Fourier
wave number $k$, which according to \cite{Kirby2005846} has more than
one conversion standards. Nevertheless, we have tested both methods
and find no significant difference between them. We show the result
that is corrected with the closed loop method in Figure
\ref{fig:ps_corr}. We find that after the correction, the power
leakage has been significantly eliminated.

\section{Discussion}
\label{sec:discussion}

\subsection{Comparison with the Polynomial Fitting Based Method}
\label{ssec:other_wt}
\begin{figure}
  \begin{center}
    \includegraphics[height=.33\textwidth,angle=270]{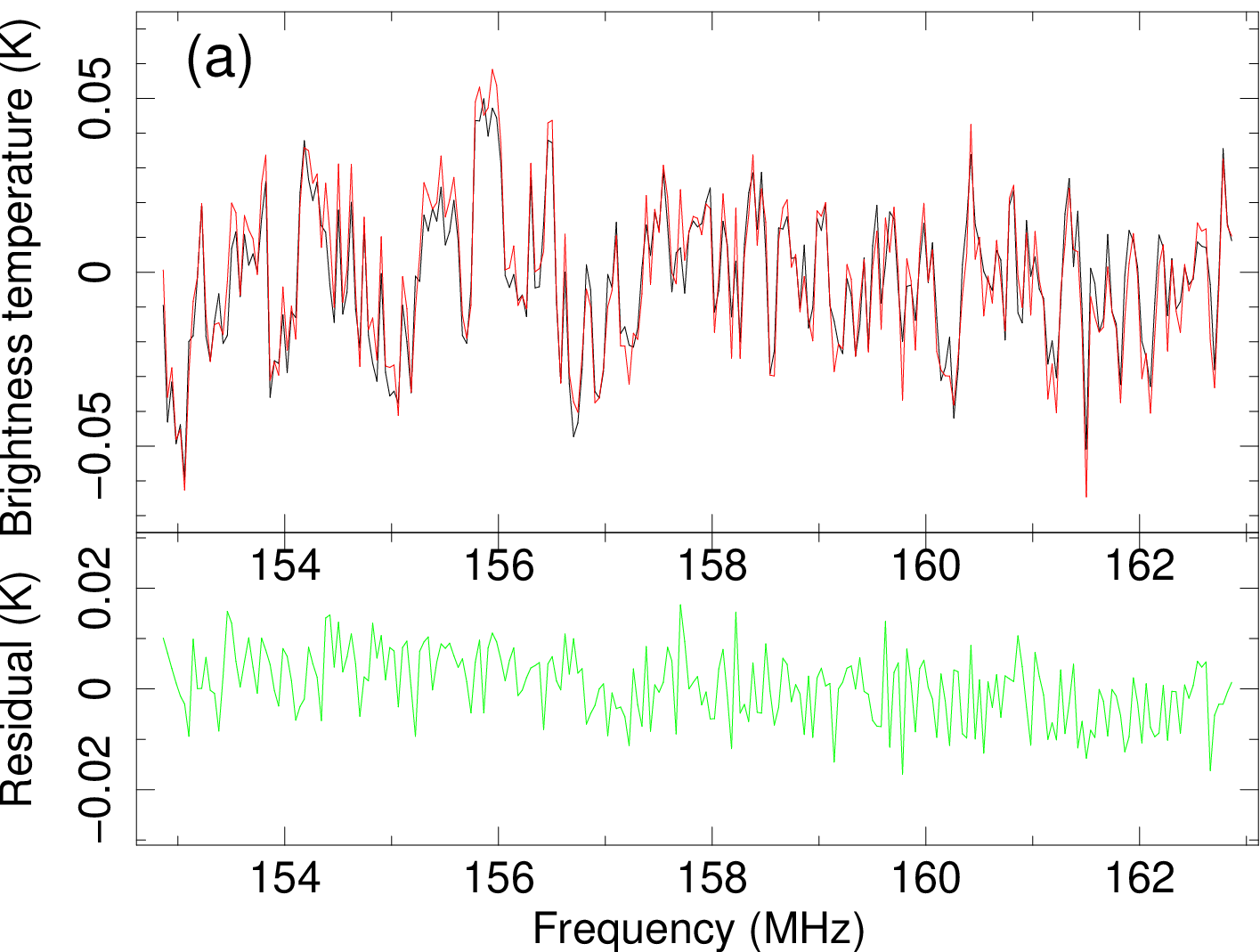}
    \includegraphics[height=.33\textwidth,angle=270]{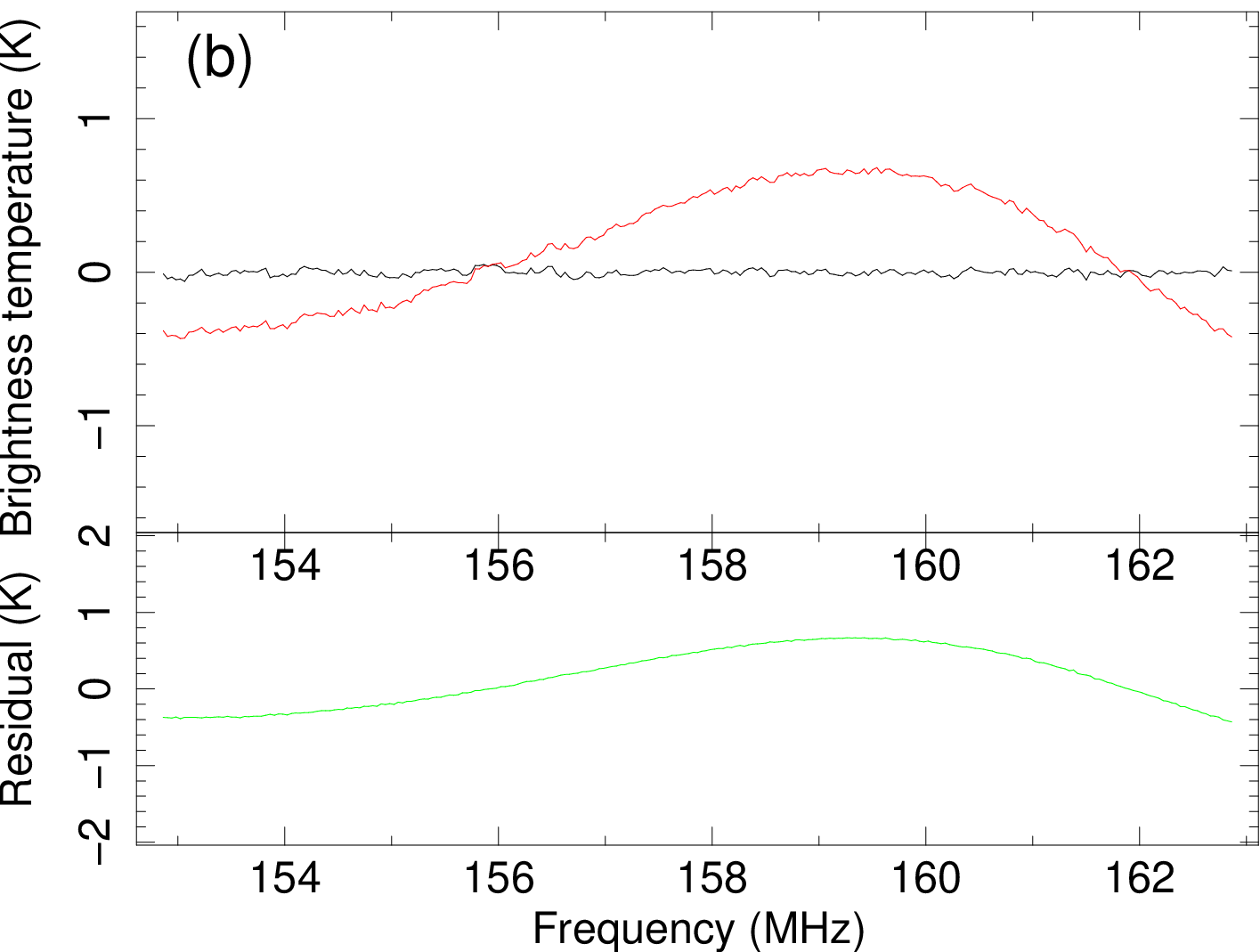}\\
    \includegraphics[height=.33\textwidth,angle=270]{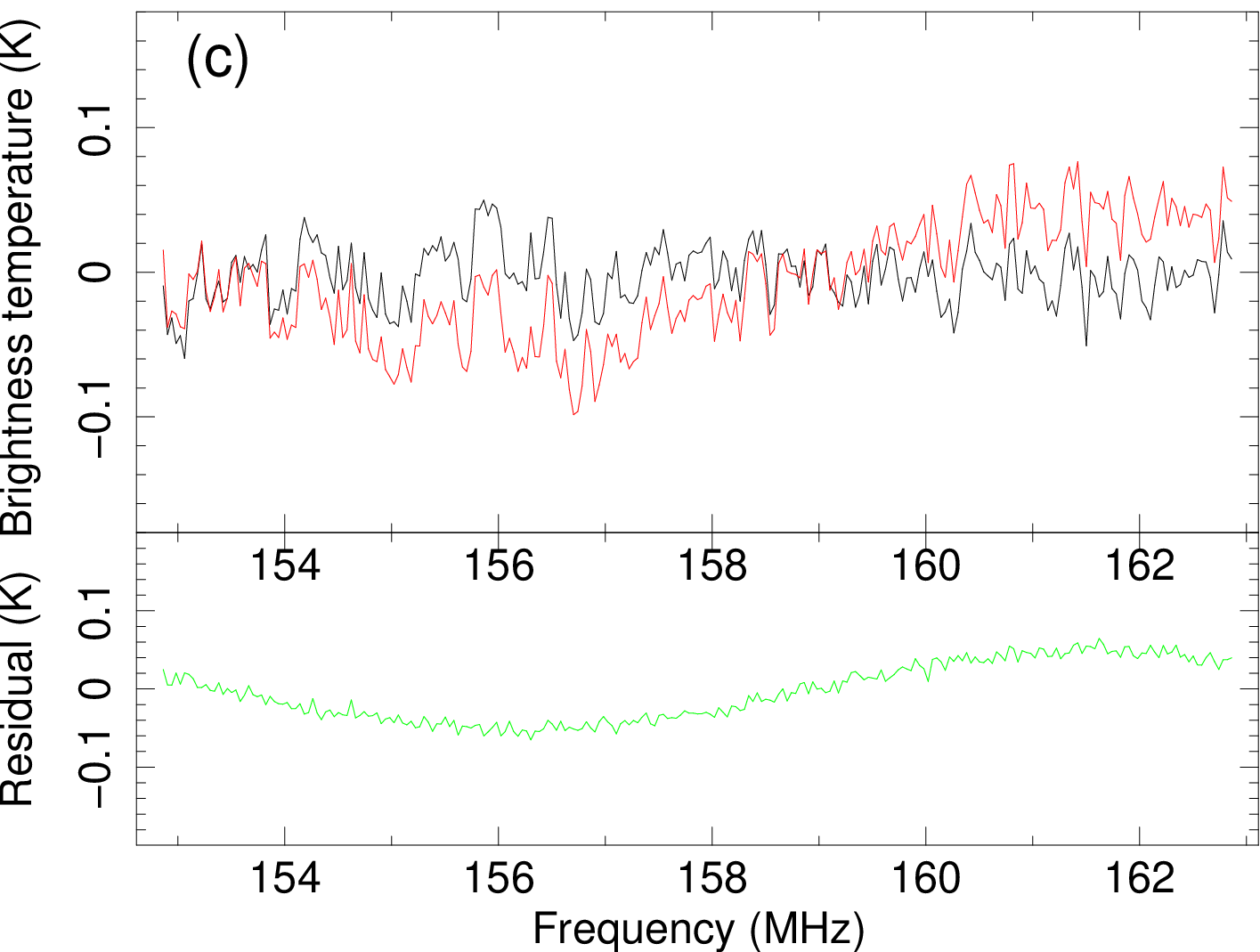}
    \includegraphics[height=.33\textwidth,angle=270]{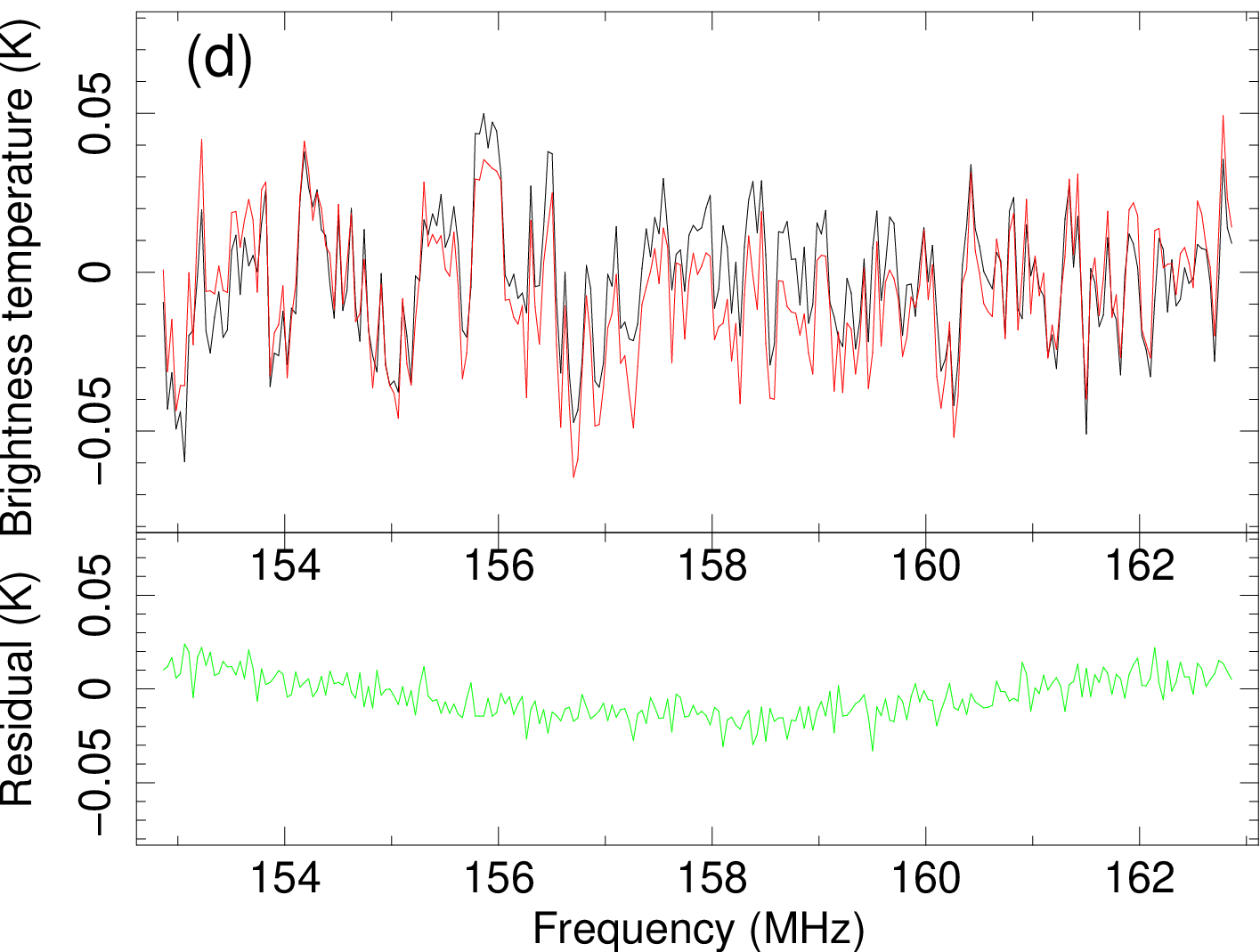}
  \end{center}
  \caption{\label{fig:result_fit} Same as Figure \ref{fig:result}, but
    the results are obtained by using the polynomial fitting based
    method.}
\end{figure}

\cite{2006ApJ...650..529W} proposed a polynomial fitting based method
for the foreground subtraction, which can be regarded as a
representative example of parametric methods. The basic idea of this
method is to fit the total spectrum with a logarithmic space $n{\rm th}$-order
polynomial
\begin{gather}
  \log T(\nu)=\sum_{m=0}^{n} a_m (\log \nu)^m,
\end{gather}
where $n$ is often chosen to be $2$ or $3$ and the residual is
regarded as the reconstructed EoR 21 cm signal. Although in our
previous work \citep{2013ApJ...763...90W} we have tested both $n=2$
and $3$ and find that $n=2$ is sufficient for FG I, we use $n=3$ here
to subtract more complex foregrounds. We present the reconstructed EoR
21 cm signal with $\Delta T_{\rm b}=6$ mK in Figure
\ref{fig:result_fit} and the estimated power spectrum of the signal
with $\Delta T_{\rm b}=60$ mK in Figure \ref{fig:ps_fit}.

For FG I, we find that both methods work, and can derive consistent
results.  For FG II, when the aim is to reconstruct the EoR 21 cm
signal in a single beam, we find that the subtraction effect of the
polynomial fitting based method in \cite{2006ApJ...650..529W} is
strongly related to $\nu_w$, while our method appears more
stable. When we only aim to estimate the one-dimensional power spectrum,
the method of \cite{2006ApJ...650..529W} works poorly for $\nu_w=5$ MHz,
and for $\nu_w$ with larger values, the estimated power spectrum is
less affected by the complexity of the foreground. For FG II with
$\nu_w=5$ MHz, the power spectral density at small $k$ is
significantly overestimated, which is obviously caused by the
contamination from the foreground. On the other hand, as has been
pointed out in our previous work \citep{2013ApJ...763...90W}, a simple
polynomial fitting method over a narrow band \citep[e.g., $2$ MHz
in][]{2006ApJ...650..529W} will also lead to the leakage of power
spectral density in the small wave number end.

\begin{figure}
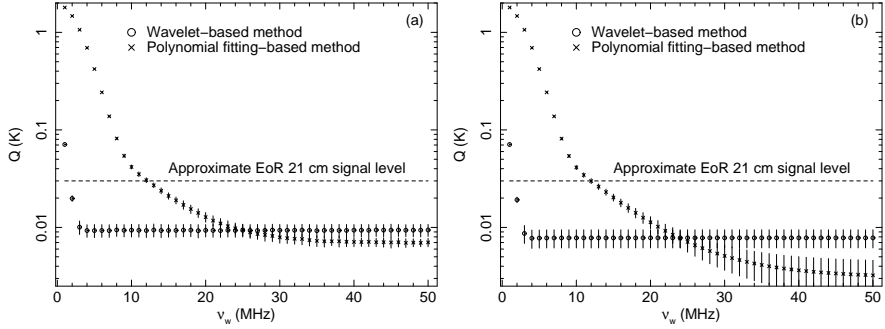

  \begin{center}
    \includegraphics[height=.35\textwidth,angle=270]{comparison1.ps}
    \includegraphics[height=.35\textwidth,angle=270]{comparison2.ps}
  \end{center}
  \caption{\label{fig:comparison} Comparison between the subtraction
    effects of the wavelet-based method and the polynomial fitting
    based method, with noise considered (a) and ignored (b).}
\end{figure}

To make a quantitative comparison, we estimate the root mean square
(rms) deviation between the input and reconstructed 21 cm signals,
which is defined as
\begin{gather}
  Q\equiv\sqrt{\frac{1}{N}\sum_{i=1}^{N}\left [T_{\rm 21cm}^\prime(\nu_i)-T_{\rm 21cm}(\nu_i)\right ]^2},
\end{gather}
where $T_{\rm 21~cm}$ and $T^{\prime}_{\rm 21~cm}$ are the input and
reconstructed 21 cm signals, respectively, and $N$ is the number of
frequency channels. A smaller $Q$ means a better subtraction
effect. Given the noise $\Delta T_{\rm b}=6$ mK, for different
$\nu_w$, we compare the RMS deviation of the results obtained with the
method of \cite{2006ApJ...650..529W} and ours as shown in Table
\ref{tbl:comp} and Figure \ref{fig:comparison}. The errors of the
estimation of $Q$ are calculated by using the standard deviation of
1000 times Monte-Carlo simulation. The comparison is performed with
the noise considered (Fig. \ref{fig:comparison}a) and ignored
(Fig. \ref{fig:comparison}b), respectively, and the result is
insensitive to the existence of noise. We find that when $\nu_w>1$
MHz, the $Q$ of our method is rather stable, and almost independent of
$\nu_w$, while the effect of the method of \cite{2006ApJ...650..529W}
seems rather sensitive to $\nu_w$. When $\nu_w<20$ MHz, our method
works significantly better than that of \cite{2006ApJ...650..529W} and
when $\nu_w>30$ MHz, the polynomial fitting based method becomes
better.

\subsection{Comparison with Wp Smoothing Method}
\cite{2009MNRAS.397.1138H} suggested a method based on the Wp
smoothing algorithm \citep[originally described
by][]{machler1993very,machler1995variational}, which is also a
non-parametric method. We implement their method according to an
implementation note written by the author i.e., {\it Implementation of
  the ``Wp'' smoothing for EoR foreground fitting} (the {\it
  Implementation note} hereafter). This method requires solving a
boundary value problem (BVP) with nonlinear terms, and when
implementing the solver numerically, it actually solves a multivariate
nonlinear system of equations. Most algorithms for solving this kind
of system of equations are based on iteration so they have the risk of
instability, and may not finally reach the optimal solution. We have
tested the Hybrid and Broyden algorithms and find that the solution is
sensitive to the initial guess.

To analyze the behavior of the iteration for solving the above BVP, we
list Equation (8)-(15) in the {\it Implementation note}. The BVP
is described as follows:
\begin{gather}
  h^\prime(x)=g(x)\\
  g^\prime(x)=p_{\mathbf{w}}(x)e^{h(x)}\left [-\frac{1}{2\lambda}\sum_{i=1}^n(x-x_i)_+\psi_i(y_i-f(x_i))\right ]\label{equ:bvp2}\\
  f^{\prime}(x)=k(x)\\
  k^{\prime}(x)=p_{\mathbf{w}}(x) e^{h(x)},\label{equ:bvp4}
\end{gather}
and the boundary conditions
\begin{gather}
  g(x_1)=0\\
  g(x_n)=0\\
  \sum_{i}\psi_i(y_i-f(x_i))=0 \label{equ:bc3}\\
  \sum_{i}x_i\psi(y_i-f(x_i))=0,\label{equ:bc4}
\end{gather}
where $\lambda$ is the Lagrange multiplier, $y_i$ is the measured
brightness temperature at frequency $x_i$, $f(x_i)$ is the solution,
which represents the smooth foreground component, and function
$\psi(x):=x\to x$. As described in the {\it Implementation note},
there is no ``natural'' value for $\lambda$, and in practice, the
authors simply choose a reasonable-looking value for $\lambda$, we set
$\lambda=1$ in our implementation. Then the recovered 21 cm signal is
obtained by $y_i-f(x_i)$.  We find that if during the iteration, the
function $h(x)$ becomes negative numbers with a relatively large
absolute value, then the right hand sides of Equations
(\ref{equ:bvp2}) and (\ref{equ:bvp4}) vanish. Then $g^{\prime}(x)$ and
$k^{\prime}(x)$ becomes zero, and finally the solution of $f(x)$ will
degenerate to a first order polynomial and $g(x)$ becomes zero. The
$f(x)$ has two free parameters i.e., the slope and the intercept,
which can be solved by Equations (\ref{equ:bc3}) and
(\ref{equ:bc4}). Obviously the following functions,
\begin{gather}
  h(x)=-C_1\label{equ:wrong_solution1}\\
  k(x)=C_2\\
  f(x)=C_2x+b\\
  g(x)=0\label{equ:wrong_solution4},
\end{gather}
where $C_1$ is a large positive number, and $C_2$ and $b$ are two
constants that can be solved with the linear equation set (Equations
(\ref{equ:bc3}) and (\ref{equ:bc4})), can be an approximate solution
to the above BVP, which however takes no information from the observed
spectrum $y_i$.

\begin{figure}
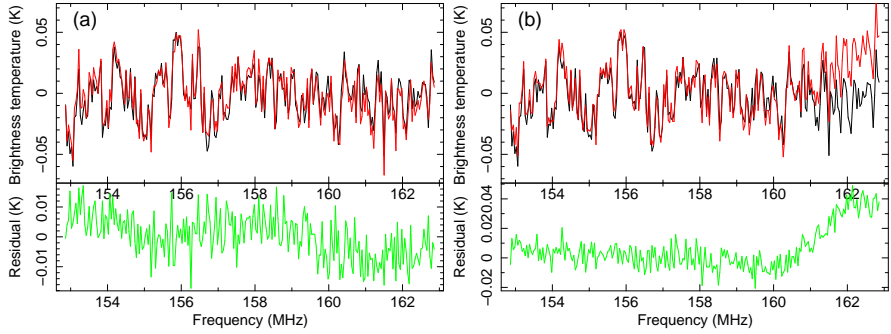

  \begin{center}
    \includegraphics[height=.35\textwidth,angle=270]{wpsmooth1.ps}
    \includegraphics[height=.35\textwidth,angle=270]{wpsmooth2.ps}
  \end{center}
  \caption{\label{fig:wpsmooth} Simulated (black), reconstructed (red)
    EoR 21 cm signals, and the residual (green) obtained with the Wp
    smoothing method by using FG I (a) and FG II with $\nu_w=5$ MHz
    (b).}
\end{figure}

In order to test the above method, we make a small modification to
prevent the iteration from reaching an obviously wrong solution, such
as Equations (\ref{equ:wrong_solution1})-(\ref{equ:wrong_solution4}).
We set a lower limit of the function $h(x)$. We test this method both
by using FG I and FG II with a $\nu_w=5$ MHz. For FG I, this method
can obtain a result as good as that of the polynomial fitting based
method, and for FG II, the $Q$ is about 0.01 K
(Fig. \ref{fig:wpsmooth}), which is close to our method. The
comparison results are summarized in Table \ref{tbl:comp}. However, we
should point out that because the solution of the BVP relies on
iteration based methods and that the size of the system of nonlinear
equations is not less than the number of channels, the process of
solution is rather time-consuming. As a rough comparison, we implement
this method by using the {\it GNU Scientific Library}
\citep{galassi2005gnu}, and run the program on a workstation with an
Intel Xeon 1.87 GHz CPU. It takes about 30 minutes to obtain the
result, while with our wavelet based method, it takes about 60 seconds
to run 1000 rounds of subtractions. For the above reason, we were not
able to calculate the errors of the estimation of $Q$ with the
Monte-Carlo method.

\subsection{Risk of Instrumental Calibration Uncertainty}
\begin{figure}
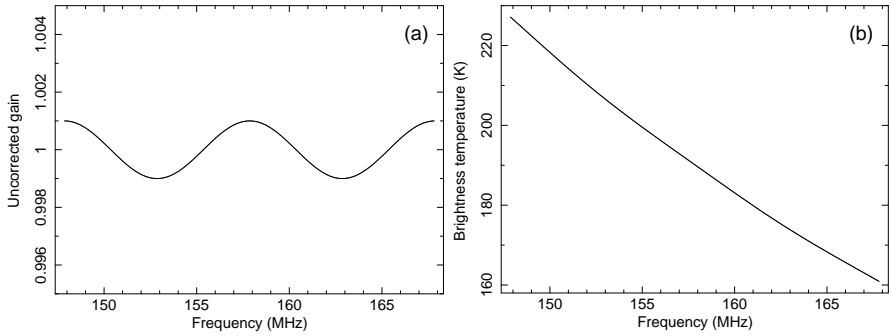

  \begin{center}
    \includegraphics[height=.35\textwidth,angle=270]{signal_gaina.ps}
    \includegraphics[height=.35\textwidth,angle=270]{signal_gainb.ps}
  \end{center}
  \caption{\label{fig:signal_gain} (a): Relative uncorrected gain
    error as a function of frequency (Equation (\ref{equ:gain})). (b): The
    total measured signal including the effect of uncorrected relative
    gain error that is shown in (a).}
\end{figure}

All above tests are based on the assumption that the instrument is
perfectly calibrated. However, calibration uncertainty, more or less,
always exists , so it is valuable to consider this effect when we test
the foreground subtraction method. As a simple test, we consider a
relative calibration error of $10^{-3}$ between different frequency
channels. We assume the uncorrected relative gain at frequency $\nu$
to be described as
\begin{gather}
  g(\nu)=1+10^{-3}\cos\left (2\pi\frac{\nu-\nu_{\min}}{10~{\rm MHz}} \right)\label{equ:gain},
\end{gather}
as shown in Figure \ref{fig:signal_gain}a. This is a rough model
composed only for the test purpose, nevertheless, according to
\cite{2010MNRAS.409.1647J}, a polarization sensitive instrument that
is improperly calibrated may possess calibration uncertainty. This
calibration uncertainty appears to be an oscillation structure along
the radio frequency axis, which is significant at several MHz scale,
just like our simple model above. In this test, we only use the above
FG I.

\begin{figure}
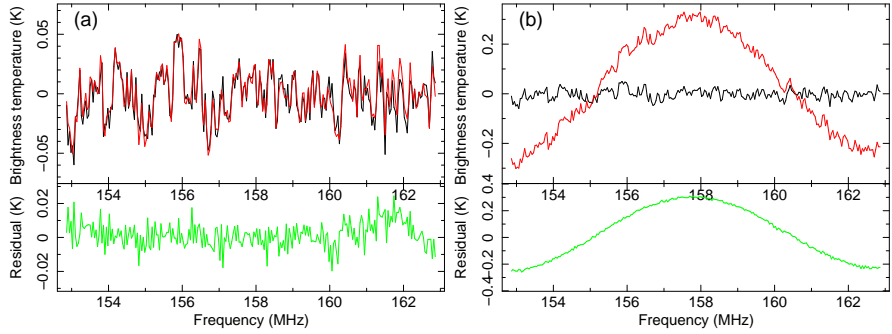

  \begin{center}
    \includegraphics[height=.35\textwidth,angle=270]{result6.ps}
    \includegraphics[height=.35\textwidth,angle=270]{result6_fit.ps}
  \end{center}
  \caption{\label{fig:calibration} Foreground subtraction result with
    the condition that a relative system calibration error of
    $10^{-3}$ is introduced. (a) The result of our wavelet based
    method. (b) The result of the polynomial fitting based method in
    \cite{2006ApJ...650..529W}. The black, red, and green lines
    represent the input, reconstructed signal, and the residual,
    respectively.}
\end{figure}

We apply the uncorrected instrumental calibration uncertainty to the
total signal (the sum of FG I, the 21 cm signal, and the instrumental
noise), and use this signal to test the subtraction method. We show
the subtraction result of the wavelet based method in Figure
\ref{fig:calibration}a. As a comparison, we test the polynomial
fitting based method, and show the result in Figure
\ref{fig:calibration}b. It's obvious that the result of the wavelet
based method is affected a little by the instrumental calibration
uncertainty, while the polynomial fitting based method becomes much
worse. This phenomenon can be explained by the fact that when the
instrumental calibration uncertainty is involved, the foreground
spectrum is no longer low-order polynomial shaped. Although we can use
a higher-order polynomial to approximate the calibration-involved
foreground, it will over-fit the background 21 cm signal. On the other
hand, the wavelet based method does not place such a strong assumption
over the foreground spectrum, i.e., polynomial-like, so the deviation
of the spectrum from a polynomial shape does not have significant
influence on the subtraction effect.

\subsection{Application to an Extremely Sharp Turnover Condition}
From above discussions we have found that the wavelet based method
appears to be more tolerant to complex conditions for the foregrounds,
and its numerical stability and computing efficiency is much higher
than the Wp smoothing based method. In this section we will test our
method with an extreme condition, i.e., FG II with $\nu_w=1$ MHz.

\begin{figure}
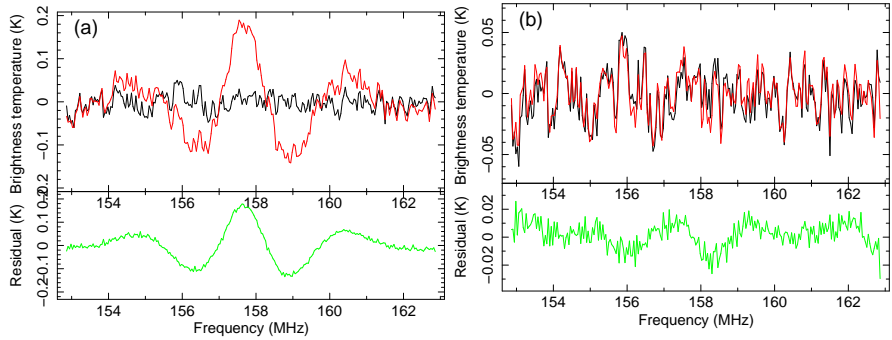

  \begin{center}
    \includegraphics[height=.35\textwidth,angle=270]{result_1MHz.ps}
    \includegraphics[height=.35\textwidth,angle=270]{man_result.ps}
  \end{center}
  \caption{\label{fig:result_1MHz} Foreground subtraction results of
    the wavelet based method with FG II with $\nu_w=1$ MHz. Panel (a)
    shows the result obtained by using the mask shown in Figure
    \ref{fig:coi}b, and panel (b) shows that obtained by using the manually
    generated mask shown in Figure \ref{fig:man_filtered}. The black,
    red, and green lines represent the input,reconstructed signal, and
    the residual, respectively.  }
\end{figure}

\begin{figure}
  \begin{center}
    \includegraphics[height=.75\textwidth,angle=270]{wt_total6.ps}\\
    \includegraphics[height=.75\textwidth,angle=270]{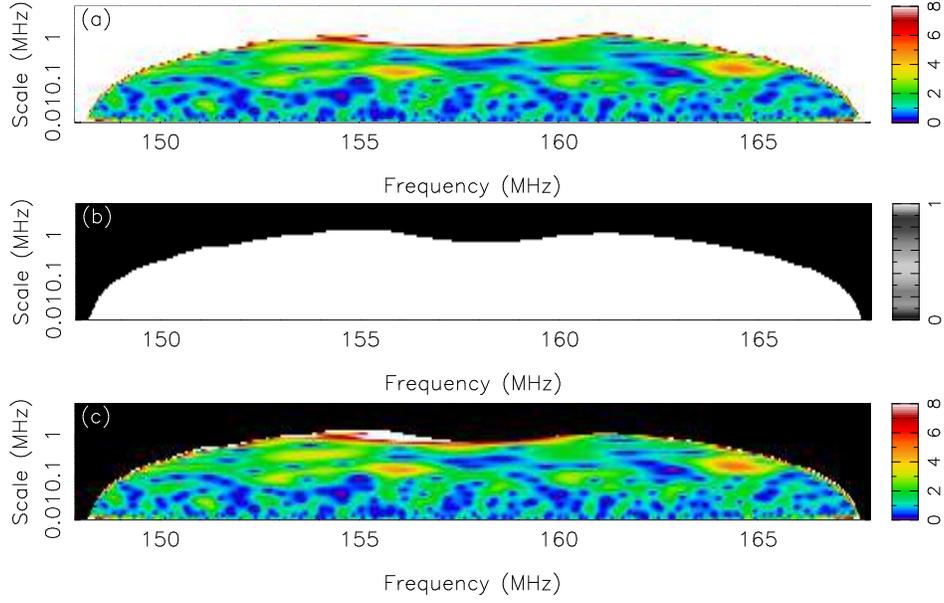}\\
    \includegraphics[height=.75\textwidth,angle=270]{wt_filtered6.ps}
  \end{center}
  \caption{\label{fig:man_filtered} (a) The absolute values of the
    wavelet coefficients of the total signal with FG II with $\nu_w=1$
    MHz.  (b) Manually generated filtering mask for FG II with
    $\nu_w=1$ MHz. (c) Corresponding filtered total signal.}
\end{figure}

By using the filtering method described in \S\ref{ssec:filtering}, we
obtain the result, which is shown in Figure
\ref{fig:result_1MHz}a. From the wavelet coefficients of the total
signal, which is shown in Figure \ref{fig:man_filtered}a, we find that
the sharp turnover of the foreground spectrum has significant
contributions to small scales, which are not filtered out by the above
filtering method. We manually draw a mask
(Fig. \ref{fig:man_filtered}b) to check whether the sharp turnover
feature can be filtered. The filtered wavelet coefficients are shown
in \ref{fig:man_filtered}c, which is transformed to real space. The
recovered EoR 21 cm signal is shown in Figure
\ref{fig:result_1MHz}b. The result has been significantly improved by
using the manual filtering method.

Although the above method is based on a subjective standard, it shows
that the wavelet based method can be further improved to handle more
complex conditions. In our future work, we will try to find out a more
objective method to subtract the foreground in such kind of extreme
conditions.

\subsection{Handling Higher and More Complex Noise}

\begin{figure}
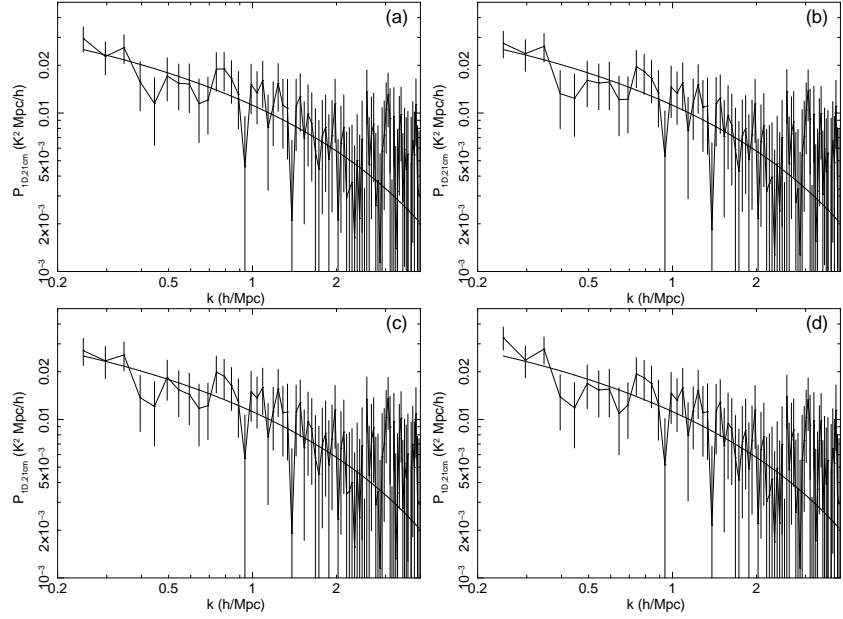

  \begin{center}
    \includegraphics[height=.33\textwidth,angle=270]{ps1_120.ps}
    \includegraphics[height=.33\textwidth,angle=270]{ps2_120.ps}\\
    \includegraphics[height=.33\textwidth,angle=270]{ps3_120.ps}
    \includegraphics[height=.33\textwidth,angle=270]{ps4_120.ps}
  \end{center}
  \caption{\label{fig:ps_120} Same as Figure \ref{fig:ps_corr}, but
    using a noise level of $\Delta T_{\rm b}=120$ mK.}
\end{figure}

We have tested two noise levels above: 6 mK for the future SKA core
region and 60 mK as a representation of current working facilities
just like what we have done in our previous work
\citep{2013ApJ...763...90W}. For the 6 mK noise level, we are able to
reconstruct actual 21 cm signal from each image pixel, while for the
60 mK noise level, we are only able to obtain the power spectrum as a
statistical information. As has been pointed in our previous work
\citep{2013ApJ...763...90W}, the 60 mK noise level is calculated based
on the 21CMA instrument, whose field of view is fixed to $5^\circ$
zone around the north celestial pole. This may not hold for other
instruments, so that in this section we test a higher noise level of
120 mK. This noise level is equivalent to reducing the total
observation time to 25\%, which may be ``more'' realistic. Still with
1000 times simulation, we obtain the one-dimensional power spectrum of
the reconstructed 21 cm signal, as shown in Figure
\ref{fig:ps_120}. We find that the results are similar to those
obtained in \S\ref{ssec:results}, but the fluctuation is larger. The
effect can be improved by increasing the number of pixels used. For
most working and upcoming facilities that are able to produce images,
the total number of pixels should be much more than 1000, so that
should be able to handle higher noise levels.

The properties of noise may appear more complex in the aspect of
stationarity. In the above tests, we assume the noise to be stationary
along both the frequency and time axes, which may be broken during
practical observations. However, as the data are accumulated before
the subtraction of foreground, the nonstationarity in time domain will
not affect our method. But what about the nonstationarity in radio
frequency domain? For a noise level significantly lower than 21 cm
signal (e.g., around several mK), this will not be a serious problem,
despite that the noise will be mixed with the 21 cm signal after the
subtraction. For a noise level significantly higher than the 21 cm
signal, the subtraction algorithm itself can still work, but more
corrections are required before producing the final result of the
power spectrum. In this condition the noise is not a white noise, so
that for excluding the power spectral density contributed by the
noise, one must subtract a more complex noise power spectrum from the
total power spectrum to produce the final result.

\subsection{Testing Other Signal Extension Methods}
\label{ssec:extension}
\begin{figure}
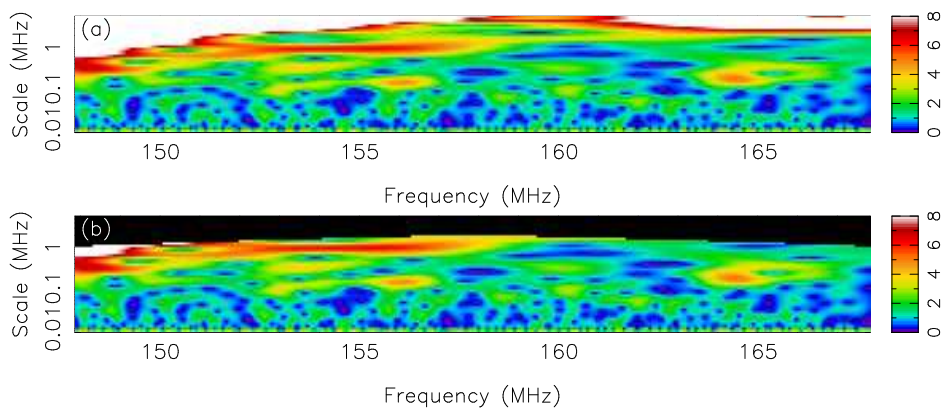

  \begin{center}
    \includegraphics[height=.75\textwidth,angle=270]{wt_total7.ps}\\
    \includegraphics[height=.75\textwidth,angle=270]{wt_filtered7.ps}  
  \end{center}
  \caption{\label{fig:pad} (a): The absolute values of the wavelet
    coefficients of the total signal with FG I calculated with the linear
    extension method (\S\ref{ssec:extension}).  (b): The absolute
    values of the corresponding filtered wavelet coefficients.}
\end{figure}

As described in \S\ref{ssec:boundary}, in the above tests, we simply
use the period signal extension method. From Figure \ref{fig:wt_fg},
we note that the significant wavelet coefficients are mainly
contributed by the boundary effect. We have also tested other
extension methods including filling zeros and symmetric
extension. We find that these extension methods differ little from
the period extension that we have used above. Nevertheless, we find that
if we extend the originally measured signal $I(\nu)$ as
\begin{gather}
  I_{\rm ext}(\nu)=\left\{
    \begin{array}{ll}
      I(\nu_{\rm min})+(\nu-\nu_{\rm min})\frac{dI(\nu)}{d\nu}|_{\nu=\nu^{+}_{\rm min}} & \nu\in[2\nu_{\rm min}-\nu_{\rm max},\nu_{\rm min})\\
      I(\nu) &\nu\in[\nu_{\rm min},\nu_{\rm max}]\\
      I(\nu_{\rm max})+(\nu-\nu_{\rm max})\frac{dI(\nu)}{d\nu}|_{\nu=\nu^{-}_{\rm max}} &\nu\in(\nu_{\rm max},2\nu_{\rm max}-\nu_{\rm min}]
    \end{array}
  \right . ,
\end{gather}
which can be named as the linear extension method, the boundary effect
can be significantly suppressed. We present the wavelet coefficients
of the total signal and the filtered signal in band $\nu_{\rm
  min}<\nu<\nu_{\rm max}$ in Figure \ref{fig:pad}. Note that the
filtering procedure is exactly the same as described in
\S\ref{ssec:filtering}, but applied to the total band after the
extension.

\begin{figure}
  \begin{center}
    \includegraphics[height=.66\textwidth,angle=270]{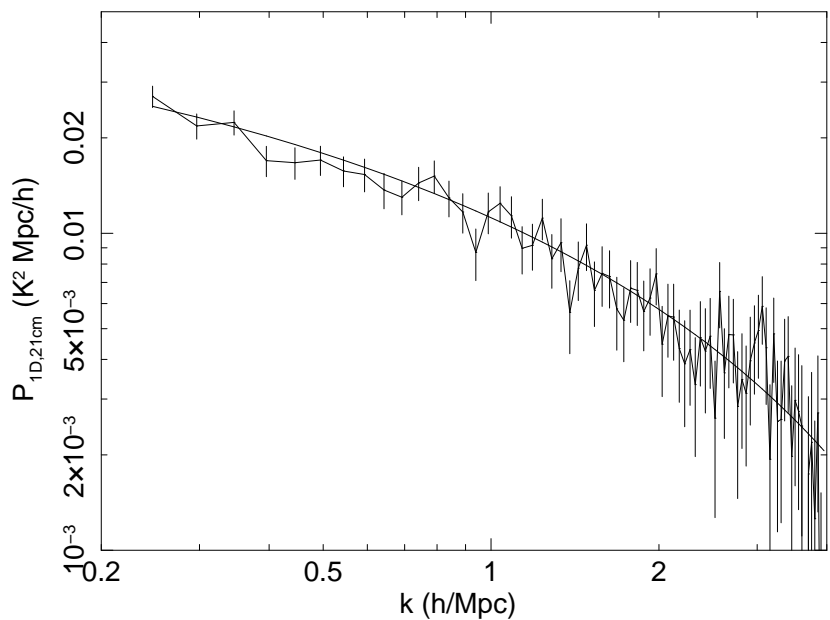}
  \end{center}
  \caption{\label{fig:ps5} Recovered one-dimensional 21 cm power spectrum with
    the linear extension method (\S\ref{ssec:extension}). Unlike
    Figure \ref{fig:ps_corr}, the power spectrum presented in this
    figure is not corrected by any method. }
\end{figure}

We roughly test the effect of foreground subtraction with this
extension method using FG I. We find that for the $\Delta T_{\rm b}=6$
mK condition, the change of $Q$ is not significant compared with the
period extension method, while for the $\Delta T_{\rm b}=60$ mK
condition, the power leakage in the small wave number end almost
disappears, as shown in Figure \ref{fig:ps5}. This apparently can be
explained as that the boundary effect mainly affects the large scale
components of the signal.

Although the simple test above show that linear extension method is a
promising method to handle the boundary effect, unlike the period
extension method that we use throughout this work, it is not commonly
used yet, and more systematic tests are required, which will be
performed in our future work. Because of the above reason, in this
work, we still use the period extension method to handle the boundary
effect.

\subsection{What Kind of Conditions are Different Methods Suitable to?}
From the discussion above, we can conclude that if the foreground
spectrum can be well approximated by a low-order polynomial, the
traditional polynomial fitting based method can work well, and obtain
an acceptable estimation of the 21 cm spectrum. When the foreground is
no longer simple, for example it appears to possess a turnover with
$\nu_w<20$ MHz; the fitting-based method will not produce an
acceptable result, but the wavelet-based method can still work
well. Furthermore, actual foreground may be more complex and can
deviate from the power-law-shaped spectrum significantly. The fitting
based method can be seriously affected. If the instrument has an
uncorrected calibration error, the wavelet based method will also have
significant advantages over the polynomial fitting based method. And
for the Wp smoothing based method, in all the conditions that we have
tested, it works at least as well as the traditional polynomial
fitting based method. When the foreground is no longer as simple as FG
I, it can obtain about the same effect as our method. However, solving
a nonlinear BVP is a rather time-consuming work, so it may be a
problem when a large number of subtraction is required, for example
when estimating power spectra.

\section{Conclusion}
\label{sec:conclusion}
We propose a CWT-based foreground subtraction method for the detection
of redshifted 21 cm signal from the EoR. This method works based on
the assumption that the foreground spectra are smooth, while the 21 cm
signal spectrum is full of saw-tooth-like structures; thus, their
characteristic scales are significantly different. We can distinguish
them in the wavelet coefficient space easily and perform the
foreground subtraction. By testing the wavelet transform based method
with a set of foreground spectra with different complexities, we find
that compared with the traditional spectral fitting based method, our
method is more tolerant to complex foregrounds. Furthermore, we also
find that when the instrument has uncorrected response errors, our
method can also work significantly better than the spectral fitting
based method. Our method can obtain similar results with the Wp
smoothing method, which is also a non-parametric method, but our
method consumes much less computing time.

\acknowledgments 

We thank the referee for his/her constructive and valuable comments,
which help improve the manuscript. This work was supported by the
Ministry of Science and Technology of China (grant Nos. 2009CB824900
and 2013CB837900), the National Science Foundation of China (grant
Nos. 11203041, 11261140641, and 11125313), the Chinese Academy of
Sciences (grant No. KJZD-EW-T01), and Science and Technology
Commission of Shanghai Municipality (grant No. 12XD1406200).

\bibliography{ms} 


\clearpage
\begin{table}
  \caption{\label{tbl:comp}The rms Deviation $Q$ of Our Method 
    and That Based on Polynomial Fitting of \cite{2006ApJ...650..529W} 
    and Wp Smoothing with $\Delta T_{\rm b}=6$ mK under Different 
    Test Conditions.}
  \begin{center}
    \begin{tabular}{c|c|c|c|c}
      \hline\hline
      \multicolumn{2}{c|}{Test Condition}&Our Method & Polynomial Fitting & Wp Smoothing$^\dag$\\
      \hline
      \multicolumn{2}{c|}{FG I}&$9.9\pm1.5$ mK&$6.8\pm0.6$ mK&$\simeq7$ mK\\
      \hline
      \multirowcell{3}[0ex][c]{FG II} & $\nu_w=5$ MHz&$9.7\pm1.4$ mK&$398\pm1.8$ mK&$\simeq10$ mK\\
      \cline{3-5}
      &$\nu_w=10$ MHz&$9.6\pm1.4$ mK&$39.1\pm2.2$ mK&-\\
      \cline{3-5}
      &$\nu_w=20$ MHz&$9.7\pm1.5$ mK&$12.5\pm1.4$ mK&-\\
      \hline\hline
    \end{tabular}
  \end{center}
  $^\dag$ Because the Wp smoothing based method is rather time-consuming, we did not perform the error estimation through the Monte-Carlo method.
\end{table}

\begin{figure}
  \begin{center}
    \includegraphics[height=.45\textwidth,angle=270]{wt_fg1.ps}
    \includegraphics[height=.45\textwidth,angle=270]{wt_fg2.ps}\\
    \includegraphics[height=.45\textwidth,angle=270]{wt_fg3.ps}
    \includegraphics[height=.45\textwidth,angle=270]{wt_fg4.ps}
  \end{center}
  \caption{\label{fig:wt_fg} (a)-(d) Absolute values of the
    wavelet coefficients of FG I and FG IIs with $\nu_w=$5, 10, and 20
    MHz, respectively.}
\end{figure}

\clearpage

\begin{figure}
  \begin{center}
    \includegraphics[height=.33\textwidth,angle=270]{ps1_fit.ps}
    \includegraphics[height=.33\textwidth,angle=270]{ps2_fit.ps}\\
    \includegraphics[height=.33\textwidth,angle=270]{ps3_fit.ps}
    \includegraphics[height=.33\textwidth,angle=270]{ps4_fit.ps}
  \end{center}
  \caption{\label{fig:ps_fit} Same as Figure \ref{fig:ps}, but the
    results are obtained by using the polynomial fitting based
    method.}
\end{figure}

\end{document}